# *Reference-free evaluation of thin films mass thickness and composition through energy dispersive x-ray spectroscopy*

Andrea Pazzaglia[1,*], Alessandro Maffini[1], David Dellasega[1], Alessio Lamperti[2], Matteo Passoni[1]

[1] Department of Energy, Politecnico di Milano, via Ponzio 34/3, I-20133 Milan, Italy

[2] Institute for Microelectronics and Microsystems, Consiglio Nazionale delle Ricerche, via Olivetti 2, I-20864 Agrate Brianza (MB), Italy

* corresponding author: andrea.pazzaglia@polimi.it (email), +39 02 2399 6354 (telephone number), +39 02 2399 3913 (fax number)
other authors emails: alessandro.maffini@polimi.it, david.dellasega@polimi.it, alessio.lamperti@mdm.imm.cnr.it, matteo.passoni@polimi.it

## Abstract

In this paper we report the development of a new method for the evaluation of thin films mass thickness and composition based on the Energy Dispersive X-Ray Spectroscopy (EDS). The method exploits the theoretical calculation of the in-depth characteristic X-ray generation distribution function ($\phi(\rho z)$) in multilayer samples, where $\phi(\rho z)$ is obtained by the numerical solution of the electron transport equation. Once the substrate composition in known, this method gives reliable measurements without the need of a reference sample and/or multiple voltage acquisitions.
The electron transport model is derived from the Boltzmann transport equation and it exploits the most updated and reliable physical parameters in order to obtain an accurate description of the phenomenon. The method for the calculation of film mass thickness and composition is validated with benchmarks from standard techniques. In addition, a model uncertainty and sensitivity analysis is carried out and it indicates that the mass thickness accuracy is of the order of 10 $\mu g/cm^2$, which is comparable to the nuclear standard techniques resolution.
We show the technique peculiarities in one example model: two-dimensional mass thickness and composition profiles are obtained for a ultra-low density, high roughness, nanostructured film.

## 1. Introduction

Measuring film mass thickness through Energy Dispersive X-Ray Sprectroscopy (EDS), also known as quantitative Electron Probe Microanalysis (EPMA), was proposed for the first time in 1960 by Sweeney, Seebold and Birks [1]. It is of a great appeal thanks to the non-destructiveness of the measurement, the use of a common apparatus (a scanning electron microscope, SEM, with an energy dispersive X-ray spectrometer, EDS), the roughness unaffected measurement and the high spatial resolution down to tens of nanometers. The technique consists of the measurement of the characteristic X-rays emitted from the sample from the atoms ionized by the primary electron beam, and then calculating the ratio (generally known as *k-ratio*) of the emitted X-ray intensities for each element in the sample from the ones emitted from homogeneous reference sample of known composition. Thus, it is possible to relate the *k-ratio* to the film mass thickness through the knowledge of the $\phi(\rho z)$ curve, which is defined as the distribution in depth of the generation of characteristic X-rays [2,3]. The knowledge of this distribution is of crucial importance in the determination of mass thickness; nevertheless its shape depends on the ionizations caused by energetic electrons, which undergo complex multiple scattering events and, consequently, a $\phi(\rho z)$ function not trivial to determine.

Since the first pioneering measurements of $\phi(\rho z)$ curves in 1951 [4] considerable effort has been expended on developing semiempirical models that properly predict the $\phi(\rho z)$ functions [5]; this resulted in the first commercial reliable software in the '90s [6-8]. The optimization of the models continues till today, and a range of software which characterize stratified samples are available [9,10]. In recent years, this technique has shown its strength also in the measurement of ultra-low density nanostructured films, which could not be characterized by most of the standard techniques [11]. Nevertheless, all this EPMA software rely on empirical or semiempirical models based on databases of measurements and so they are limited by a low flexibility and by the intrinsic uncertainty of the experimental data. In addition, most of them suffer from methodological limitations [12]: the measurement procedure which assures the highest accuracy needs EDS measurements of a standard reference sample having the same composition of the substrate or the film, which is not always available, and it needs also many acquisitions, higher than 3, at different accelerating voltages, which involves time-consuming measurements.

Only recently, an attempt to obtain $\phi(\rho z)$ curves in a theoretical way has been carried out. Using a Monte Carlo approach it is possible to simulate the electron transport phenomenon also in complex geometries [13-16] and consequently also the X-ray generation distributions can be calculated, in order to obtain measurements of film mass thickness [17]. However, it is well known that Monte Carlo methods suffer from statistical errors that can be reduced only with high computational time, which goes in the direction of undesirable long procedures for the purpose of thin film mass thickness measurements. In addition, a recent reliability analysis showed systematic discrepancies between the simulated and the experimental values of *k-ratios* [18].

On the other hand, the problem of electron transport in solids can be dealt using a kinetic approach, using the Boltzmann transport equation: exploiting some reasonable assumptions and the system symmetry, a simpler equation, in the Fokker Planck form, was derived, which can easily be solved numerically [19,20]. This powerful equation was used, in a stationary form, to calculate the deposition of energy and the fraction of backscattered electrons in homogeneous thin films [21,22]. This transport equation, in the time-dependent form, allowed also the calculation of the $\phi(\rho z)$ function in the case of homogeneous samples [23,24]. Although this approach could lead to several advantages, as the theoretical determination of $\phi(\rho z)$ and the quite easy numerical solution, it was not adopted in the determination of the X-rays generation distribution in multilayer geometries. In addition, the lack of precision in the involved physical parameters resulted in a significant inaccuracy in the $\phi(\rho z)$ function.

This work lies within this particular approach, and we present important improvements: the solution of the electron transport equation is carried out with more precise physical parameters and in a multilayer geometry. The resultant $\phi(\rho z)$ functions are consequently more reliable and, thanks to the multilayer description, they enable relating thin film mass thickness and composition to EDS data using an innovative method: instead of the k-ratios, the ratio of the film X-rays intensities over the substrate intensity overcomes the need of reference samples and multiple voltages measurements, once the substrate composition is known. Reference-free measurements are particularly appealing in the material science, where thin films of unknown composition and mass thickness are often deposited into well characterized substrates.

## 2. Reference-free mass thickness and composition evaluation

The aim of this work is to retrieve thin film mass thicknesses $\tau$ and compositions $C_{F,k}$ in a film-substrate geometry, from EDS measurements, without the need of reference samples and multiple voltages measurements. In the literature, the mass thickness determination is done through the measurements of the ratio of characteristic X-rays emitted from the multilayer sample with respect to a reference homogeneous sample, known as *k-ratios* ($k = I_F/I_F^{ref}$ or $k = I_S/I_S^{ref}$, where the $F$ and $S$ subscripts refer to the film and substrate layers respectively), and from the knowledge of the in depth X-ray generation distributions $\phi(\rho z)$. In this work we propose using a different approach based on the measurement of different ratios, called *K-ratios* (with a capital 'K' to distinguish it from the conventional *k-ratio*), of film X-ray intensities over the substrate ones which enables, when the substrate composition $C_{S,j}$ is known (which is a common situation), to overcome the need of reference samples. The method consists of relating the *K-ratios* to the thin film mass thickness $\tau$ and composition $C_{F,k}$ (see Fig. 1a). This is done through the following equation:

$$K_{k,j} = \frac{I_{F,k}}{I_{S,j}} = \frac{\varepsilon_k C_{F,k} \int_0^\tau \phi_{F,k}(\rho z) \exp(-\chi_{F,k}\rho z) \, d(\rho z)}{\varepsilon_j C_{S,j} \exp(-\chi_{F,j}\tau) \int_\tau^\infty \phi_{S,j}(\rho z) \exp(-\chi_{S,j}\rho z) \, d(\rho z)} \quad (1)$$

Where $I_{F,i}$ is the intensity of measured X-rays generated in the film by the $k$-th element while $I_{S,j}$ is the substrate intensity of the $j$-th element; $\varepsilon_k$ and $\varepsilon_j$ are the detector efficiencies at the X-ray energies of the $k$-th and $j$-th elements respectively; $C_{F,k}$ and $C_{S,j}$ are the atomic fractions of $k$-th and $j$-th elements in the film and in the substrate, $\chi = \mu/(\rho \cos\alpha)$ is the attenuation coefficient ($\alpha$ is the angle of the detector with respect to the interface of the sample) and the subscripts refer respectively to the layer and the X-ray energy.

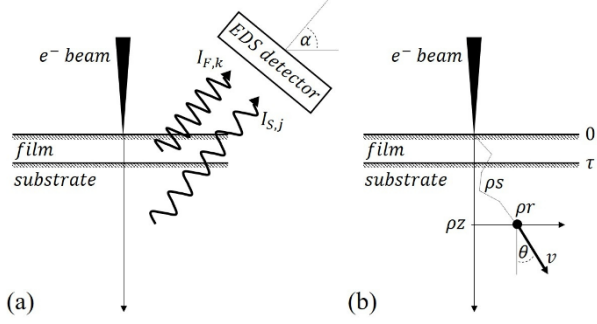

***Fig. 1.*** *Scheme of the problem geometry. In (a) the characteristic X-rays emission and measurement is represented, while in (b) a sample electron trajectory is shown, with the variables that are used in the analytical treatment of the electron transport.*

Thus, if the $\phi(\rho z)$ distributions are known, the equation (1) can be numerically solved to obtain, from measured K-ratios, the film mass thickness $\tau$ and the film atomic composition $C_{F,k}$; the calculation of the X-ray generation as a function of depth is not straightforward because it depends on the complex physics of electrons multiple scattering with the additional problem of the multilayer geometry, which can introduce strong perturbations to the $\phi(\rho z)$ curves compared to a homogeneous bulk sample.

Consequently a theoretical approach should be used to describe the electron transport into multilayer samples and then to derive accurate $\phi(\rho z)$ functions; in particular we adopt a kinetic approach which is described in Section 2.1.

We also point out that the relation (1) is based on the spatial one dimensional approximation, namely that all the quantities depend only on the sample depth variable, which is valid in the cases where $\tau$ and $C_{F,k}$ vary slowly with respect to the lateral distribution of X-ray generation $\psi(\rho r)$; we calculate this function in Section 2.2, with a hybrid fluid-kinetic approach, in order to retrieve the EDS measurements lateral resolution (1).

## 2.1. Electron transport model

The X-ray generation depth distribution $\phi(\rho z)$ can be calculated with a kinetic approach from the knowledge of the electron distribution function $F(\vec{r}, \vec{p}, t)$, which solves the Boltzmann transport equation. We exploit the problem symmetries, by reference to Fig. 1b, to neglect the lateral spatial coordinates $x$ and $y$; then, if we express the momentum by the energy and the orientation, within the spherical coordinates, we can also neglect the azimuthal angle $\varphi$. In addition, it is useful to relate the time to the path travelled by an electron via its velocity. In these new coordinates the number of variables of the electron distribution function are thus reduced from 7 to 4: one spatial dimension, two momentum dimensions and one time variable, $F(\vec{r}, \vec{p}, t) \to F(z, \vartheta, E, vt = s)$, where $z$ is the depth, $\vartheta$ the angle of the velocity vector with respect to the $z$ axis, $E$ the energy and $s$ the path length travelled by the electron. Finally, it is more convenient to express the spatial variables in terms of mass and we obtain $F(\rho z, \vartheta, E, \rho s)$. Accordingly, the $\phi(\rho z)$ function is calculated as an integration of the distribution function over all the variables except $\rho z$, with a weight function which represents the microscopic ionization cross section $\sigma_{ion}$, multiplied by the fluorescence yield $\omega$, the atomic fraction $C$ and the atoms number density, namely the number of atoms per unit volume:

$$\phi_{F/S,i}(\rho z) = \frac{\omega_{F/S,i} C_{F/S,i} N_{av}}{A_{F/S,i}} \int_0^{eV} \int_0^{\rho s_R} \int_0^{\pi} F(\rho z, \theta, E, \rho s) \sigma_{F/S,i}^{ion}(E) \sin\theta d\theta d(\rho s) dE \quad (2)$$

Where $N_{av}$ is the Avogadro number and $A$ the atomic weight (in $mg/mol$ units), $eV$ is the initial electron energy, which corresponds to the accelerating voltage times the electron charge, while $\rho s_R$ is the electron mass range (calculated using an empirical relation or the stopping power integration [25]) in $mg/cm^2$ units. Then $\phi(\rho z)$ has the units of $1/(mg/cm^2)$ and, considering that each energetic electron carries $e$ charge, we can also normalize the distribution to the electron gun current and express it in $1/(\mu A\, mg/cm^2)$ units.

In order to make feasible the determination of $F$ via the Boltzmann transport equation, some useful assumptions can be exploited, in order to further simplify the equation:

1. The electrons collide with atoms in two decoupled ways: elastic and inelastic collisions
2. The elastic collisions change the electron trajectories without affecting the electron energies (we neglect the atom recoil energy)

3. The inelastic collisions make the electrons lose energy without affecting their trajectories (the momentum transfer is negligible with respect to elastic collisions)

These assumptions are reasonable (for a detailed justification see [26]) and they enable to separate the electron distribution function:

$$F(\rho z, \vartheta, E, \rho s) = f(\rho z, \vartheta, \rho s) g(E, \rho s) \tag{3}$$

Including this relation in the Boltzmann equation and applying the variable separation method we obtain two coupled transport equations:

$$\frac{\partial f}{\partial(\rho s)} = -\vec{u} \cdot \nabla_{\rho r} f + \left(\frac{\partial f}{\partial(\rho s)}\right)_{coll} \tag{4}$$

$$\frac{\partial g}{\partial(\rho s)} = \left(\frac{\partial g}{\partial(\rho s)}\right)_{coll} \tag{5}$$

Where $\vec{u} = \vec{v}/v$ and $\nabla_{\rho r} = \frac{\partial}{\partial \rho x}\hat{i} + \frac{\partial}{\partial \rho y}\hat{j} + \frac{\partial}{\partial \rho z}\hat{k}$ is gradient operator with respect to the mass coordinates. From equation (4), expressing the integral of collisions and expanding in power series [19], an equation which describes the elastic multiple scattering process can be derived:

$$\frac{\partial f(\rho z, \theta, \rho s)}{\partial(\rho s)} = -\cos\theta \frac{\partial f(\rho z, \theta, \rho s)}{\partial(\rho z)} + \frac{1}{\sin\theta}\frac{\partial}{\partial\theta}\left(\sin\theta \frac{\partial}{\partial\theta}\left(\frac{f(\rho z, \theta, \rho s)}{\rho\lambda_{tr}(\rho z, \bar{E})}\right)\right) \tag{6}$$

Where $\rho\lambda_{tr}$ is the mass transport mean free path, which is a functional of the differential elastic cross section $\partial\sigma_e(\theta, E)/\partial\theta$:

$$\frac{1}{\rho\lambda_{tr}}(E) = \frac{N_{av}}{A_r}\int \pi(1-\cos\theta)\frac{\partial\sigma_e(\theta, E)}{\partial\theta}\sin\theta \, d\theta \tag{7}$$

The dependence of the transport mean free mass path on the depth variable is due to the multilayer geometry, so this parameter follows a piecewise trend along $\rho z$. It should be noted that the coupling between the two transport equation is mediated by the $\rho\lambda_{tr}$ parameter because it is evaluated at the mean energy $\bar{E} = \int E \, g(E, \rho s) \, dE$ which is calculated from the solution of equation (5), which describes the electron energy loss process.

The energy transport problem could be solved with the continuous slowing down approximation (CSDA), which enables to express the energy with a one-to-one relation to the path length; however this treatment oversimplifies the problem, where the electron energy spectrum broadens after few inelastic collisions and consequently the energy straggling plays an important role. Thus, in order to take into account the energy straggling we can express equation (5) with the following relation:

$$\frac{\partial g(E, \rho s)}{\partial(\rho s)} = -g(E, \rho s)\int_0^E \frac{\partial\sigma_i(E, W)}{\partial W}dW + \int_0^{eV-E}\frac{\partial\sigma_i(E+W, W)}{\partial W}g(E+W, \rho s)dW \tag{8}$$

Where V is the initial electron energy, while $\partial\sigma_i(E, W)/\partial W$ is the differential inelastic scattering cross section which expresses the probability of an electron of energy E to lose energy W in an inelastic collision. In the literature several approaches have been proposed to describe the straggling distribution as a function of the path travelled by charged particles, as the Gaussian model or the Landau distribution [27]. These models, however, suffer from some limitations, as the low accuracy at high travelled path and the difficulty in including the multilayer geometry in the calculation; for this reason we decided to adopt the convolution method which is widely regarded as the most accurate approach to calculate the exact theoretical straggling distribution, being only limited by the accuracy of the numerical solution and the uncertainty on the knowledge of the differential inelastic scattering cross section.

The convolution method consists in calculating via a numerical method the energy spectrum after a given path $\Delta(\rho s)$, as the sum of the energy loss spectra due to $n_{ic}$ inelastic collisions, multiplied for the related probability, given by the Poisson distribution.

In detail, the treatment is based on the calculation of the mass inelastic mean free path:

$$\frac{1}{\rho \lambda_{ic}}(\bar{E}) = \frac{N_{av}}{A_r} \int_0^{\bar{E}} \frac{\partial \sigma_i(\bar{E}, W)}{\partial W} dW \tag{9}$$

And then calculating the mean number of inelastic collision in a given mass path length as $n_{ic} = \Delta(\rho s)/\rho \lambda_{ic}$. Thus, starting from a initial monoenergetic distribution, after $n_{ic}$ collisions the energy spectrum will be given by the $n_{ic}$-fold convolution of the differential inelastic cross section:

$$w^0(E) = \delta(E - eV)$$
$$w^{k+1}(E) = \int_0^E \frac{\partial \sigma_i(E, W)}{\partial W} w^k(E + W) dW \tag{10}$$

And taking into account also the statistics of collision number, namely the Poisson distribution, we finally obtain the equation that describes the electron energy straggling:

$$g(\rho s + \Delta(\rho s), E) = \int_0^E g(\rho s, E') \sum_{k=0}^{\infty} \frac{n_{ic}^k e^{-n_{ic}}}{k!} w^k(E') dE' \tag{11}$$

It should be noted that this equation is coupled with the spatial transport equation (6) because of the multilayer geometry, which is addressed in detail in Section 2.3.

## 2.2. Spatial advection-diffusion model

As stated in the first part of Section 2.1, the described transport model enables to calculate the electron distribution function only on one spatial dimension, the sample depth; nevertheless, in order to evaluate the lateral EDS resolution, it is of high interest to retrieve information about the electron radial distribution and, consequently, about the X-ray generation radial distribution.

This is done by solving another differential equation which is obtained by a hybrid fluid-kinetic approach, considering the electrons as a fluid in cylindrical coordinates, with advection and diffusion coefficients, variable in space, which are retrieved from the $f$ and $g$ distributions of the kinetic approach.

Accordingly, the continuity equation, $\partial n/\partial t = -\nabla \cdot \vec{j}$, where $n$ is the electron spatial distribution along the depth $z$ and the radius $r$ and $\vec{j}$ is the electron current, can be expressed in cylindrical coordinates as $\partial n/\partial t = -\partial j_z/\partial z - 1/r \, \partial/\partial r \, (r \, j_r)$.

If we take into account the fact that the currents depend on time and space and we change the time coordinate into the mass path $\rho s$ and the spatial coordinates into the mass spatial coordinates $\rho z$ and $\rho r$, we obtain:

$$\frac{\partial n(\rho z, \rho r, \rho s)}{\partial (\rho s)} = -\frac{\partial j_z(\rho z, \rho r, \rho s)}{\partial (\rho z)} - \frac{1}{\rho r} \frac{\partial}{\partial (\rho r)} \rho r \, j_r(\rho z, \rho r, \rho s) \tag{12}$$

Where $n$ is the electron spatial distribution along the depth and the mass radius $\rho r$ and $\vec{j}$ is the electron current which is spatial and path dependent.

The problem is then to retrieve $j_z$ and $j_r$ from the knowledge of $f$. The net current along the depth direction depends on the angular distribution of electrons at a given depth and is derived as the projection of the distribution along the mass depth axis:

$$j_z(\rho z, \rho r, \rho s) = n(\rho z, \rho r, \rho s) \int_0^{\pi} f(\rho z, \theta, \rho s) \cos\theta d\theta \tag{13}$$

In this treatment we should also take into account the azimuthal angle distribution, which is uniform, in the calculation of the projection integral, however we neglect it for the sake of simplicity and also because the final result is not highly affected. It should be noted that in this treatment $a(\rho z) = \int_0^{\pi} f(\rho z, \theta, \rho s) \cos\theta d\theta$ acts as a depth-dependent advection term. On the other hand, the radial electron current is not due to advection, because the angular distribution is symmetric and $\int_{-\pi}^{\pi} f(\rho z, \theta, \rho s) \sin\theta d\theta$ is equal to 0; consequently, the radial current is due to diffusion and is calculated by:

$$j_r(\rho z, \rho r, \rho s) = -dr \int_0^{\pi} f(\rho z, \theta, \rho s) \sin\theta d\theta \, \frac{\partial n(\rho z, \rho r, \rho s)}{\partial (\rho r)} \tag{14}$$

The term $dr \int_0^\pi f(\rho z, \theta, \rho s) \sin\theta d\theta = D(\rho z)$ acts as a diffusion factor in the final form of the equation:

$$\frac{\partial n(\rho z, \rho r, \rho s)}{\partial(\rho s)} = -\frac{\partial}{\partial(\rho z)} a(\rho z) n(\rho z, \rho r, \rho s) + D(\rho z) \frac{1}{\rho r} \frac{\partial}{\partial(\rho r)} \rho r \frac{\partial}{\partial(\rho r)} n(\rho z, \rho r, \rho s) \qquad (15)$$

Once this advection-diffusion equation is solved it is possible to retrieve the radial distribution of X-rays generation which is the function needed for the estimation of the lateral spatial resolution of EDS measurements. Similarly to the $\phi(\rho z)$, it is obtained by an integration over all the variables except the radial one, with the ionization cross section multiplied for the fluorescence yield as the weight function:

$$\psi(\rho r) = \frac{\omega N_{av}}{A_r} \int_0^{eV} \int_0^{\rho s_R} \int_0^{\rho s_R} n(\rho z, \rho r, \rho s) g(E, \rho s) \Sigma_i(E) d(\rho z) d(\rho s) dE \qquad (16)$$

Consequently, the radial resolution $\widetilde{\rho r}$ can be defined as the radius at which the X-ray intensity is reduced by a factor e, namely $\psi(\widetilde{\rho r}) = \psi(0)/e$. The calculated resolution is a crucial parameter in the characterization of films with high roughness or with varying composition, in fact the transport model lye on the hypothesis of $\tau$ and $C_{F,k}$ varying slowly with respect to the lateral resolution. Consequently the condition $\widetilde{\rho r}/\tau \cdot \partial\tau/\partial\rho r \ll 1$ must be fulfilled.

## 2.3. Numerical solution

The equation (6) and (11) are coupled and their solution is carried out numerically at the same time. The equation (6) is solved with an explicit finite difference scheme: a first order accuracy upwind scheme on the ρz axis and a centered second order accuracy scheme on the θ axis [28]. The scheme is solved onto the grid $\rho z \times \theta$, with 80 x 40 cells over the ranges $[0, \rho s_R] \times [0, \pi]$, with the following initial and boundary conditions:

$$\begin{aligned} f(\rho z, \theta, \rho s) &= \delta(\rho z)\delta(\theta) \ at \ \rho s = 0 \\ f(\rho z, \theta, \rho s) &= 0 \ at \ \rho z = 0 \ and \ 0 \leq \theta < \pi/2 \ and \ \rho s > 0 \\ f(\rho z, \theta, \rho s) &= 0 \ at \ \rho z = \rho s_R \\ \frac{\partial f(\rho z, \theta, \rho s)}{\partial \theta} &= 0 \ at \ \theta = 0, \pi \end{aligned} \qquad (17)$$

The first condition is the initial one, where electrons are all at the surface point without angular dispersion; the meaning of the second condition is that there is not injection of new electrons at the surface after the initial 'time' (when $\rho s = 0$), while it is possible to have electrons escaping from the surface at any 'time' ($f(\rho z, \theta, \rho s) \neq 0$ at $\rho z = 0$ and $\pi/2 < \theta \leq \pi$ and $\rho s > 0$); the third condition imposes that the electron distribution function vanishes at the electron range and the fourth condition expresses that the net distribution angular flux is null at the boundaries, because of the angular symmetry.

In order to assure stability to the method, some shrewdness must be carried out. In fact, the electron flow direction changes in the θ axis in correspondence to the $\pi/2$ value and so also the direction of the finite derivative in the ρz axis (first term on the right hand side of equation (6)) must change at that grid line.

In addition, it should be noticed that the second term of the right hand side of equation (6) plays the role of a diffusion term with a not-constant coefficient $1/\rho\lambda_{tr}(\rho z, \overline{E})$ in 'time' and space, which increases dramatically when the energy decreases, namely when ρs increases; to ensure the numerical scheme stability, the Courant-Friedrichs-Lewy condition must be fulfilled at each step, namely introducing a variable $\Delta(\rho s)$ step which decreases with the path length: $\Delta(\rho s) < (1/\Delta(\rho z) + 2/(\rho\lambda_{tr})_{max}\Delta\theta^2)^{-1}$ [28].

The solution in a multilayer geometry is accounted by changing with a piecewise function the value of the $1/\rho\lambda_{tr}(\rho z, \overline{E})$ coefficient along the ρz axis. Nevertheless, if the jump discontinuity is too high, some instabilities can arise; for this reason we have expressed all the variables as the mass ones, in a way that the differences of film and subtrate densities do not contribute to this discontinuity and make the solution method stable.

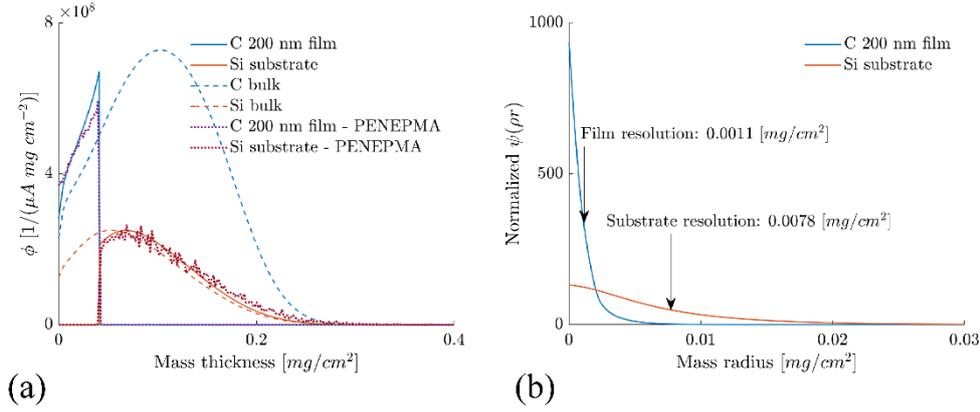

**Fig. 2.** *Figure (a) shows the calculated $\phi(\rho z)$ distributions by the electron transport model and by PENENPMA (with $2 \cdot 10^6$ trajectories) at an accelerating voltage of $10\ kV$, in 3 different geometries: a single semi-infinite layer of bulk Carbon by our model (blue dashed curve), a single semi-infinite layer of bulk Silicon by our model (orange dashed curve), a film-substrate geometry made by a bulk density, 200 nm thick, Carbon film on a semi-infinite Silicon substrate by our model (blue and orange continuous curves) and by PENENPMA (violet and red points curves). It should be noted that in the film-substrate geometry the $\phi(\rho z)$ functions are perturbed with respect to the single semi-infinite samples. In addition, the distribution calculated with our model are in good agreement with ones calculated with the Monte Carlo code. Figure (b) shows the calculated $\psi(\rho r)$ distributions by the advection-diffusion equation at an accelerating voltage of 10 kV, with 20 nm electron beam diameter, in a film-substrate geometry made by a bulk density, $200\ nm$ thick, Carbon film and semi-infinite Silicon substrate. The radial resolution values, $\widetilde{\rho r}$, are reported on the graph.*

The numerical solution of equation (11) is straightforward and is carried out over the energy grid, uniformly spaced by $0.05\ keV$. However it should be noted that in the film-substrate geometry we have to distinguish the energy distributions in the different layers, because of their different energy loss, and we have to evaluate the $g$ functions for each layer, namely $g_F$ in the film and $g_S$ in the substrate; in addition, it must be also considered that the flow of electrons from one layer to another one tend to mix the energy spectra. This phenomenon is taken into account calculating the fraction of electrons, $r_F$ and $r_S$, respectively coming from the film to the substrate and from the substrate to the film at each path step:

$$r_F(\rho s) = \frac{\int_0^{\frac{\pi}{2}} f(\tau, \theta, \rho s)\sin\theta\cos\theta d\theta\ d(\rho z)}{\int_\tau^{\rho s_R} \int_0^\pi f(\rho z, \theta, \rho s)\sin\theta d\theta d(\rho z)}$$
$$r_S(\rho s) = \frac{\int_{\frac{\pi}{2}}^{\pi} f(\tau, \theta, \rho s)\sin\theta\cos\theta d\theta d(\rho z)}{\int_0^\tau \int_0^\pi f(\rho z, \theta, \rho s)\sin\theta d\theta d(\rho z)} \tag{18}$$

We first calculate the unperturbed energy distributions in the film and the substrate, $g^0_{S/F}$; then we take into account the mixing phenomenon by averaging the film and substrate distributions by the fractions $r_F$ and $r_S$:

$$g_F(\rho s, E) = \big(1 - r_F(\rho s)\big) g_F^0(\rho s, E) + r_S(\rho s)g_S^0(\rho s, E)$$
$$g_S(\rho s, E) = \big(1 - r_S(\rho s)\big) g_S^0(\rho s, E) + r_F(\rho s)g_F^0(\rho s, E) \tag{19}$$

Once the electron distribution, $F = fg$, is calculated, it is possible to calculate the $\phi(\rho z)$ functions for both the film and the substrate with equation (2), remembering that also the microscopic ionization cross section and the fluorescence yield are described by piecewise functions.

As we can see from the example in Fig. 2a, the model enables to calculate the $\phi(\rho z)$ functions in a multilayer geometry with a good agreement with respect to the PENEPMA Monte Carlo code, with two main advantages, namely the faster calculation (about one minute for our model against about two hours for the Monte Carlo code) and the smooth feature of the analytical curve. In addition, it is possible to observe the perturbation in the distributions due to the multilayer geometry with respect to the single layer case. For example, in the shown case, the electron transport features change because of the jump discontinuity in the $1/\rho\lambda_{tr}(\rho z, \bar{E})$ coefficient at the $\tau$ interface; the diffusion coefficient is higher in the substrate than in the film, then the electrons tend to be more backscattered by the substrate toward the film, and thus the film electron density and $\phi(\rho z)$ distribution increases, with respect to the single layer case.

Finally, once the film mass thickness and composition is fixed, the knowledge of the $\phi(\rho z)$ functions in the film-substrate geometry can be used to simulate the values of the $K_{k,j}$ ratios; the problem is then to minimize the difference between the calculated ratios with the measured ones by EDS, $\widetilde{K_{k,j}}$, namely to minimize the chi squared factors:

$$\chi^2_{k,j} = \left(K_{k,j} - \widetilde{K_{k,j}}\right)^2 \quad (20)$$

Two different algorithm can be used: the first one is based on the gradient descent algorithm [29] which consists of calculating the gradients of the simulated $K_{k,j}$ ratios with respect to the mass thickness and composition and then to obtain, with a linear regression, the next-iteration values of mass thickness and composition; with few iterations this method enables to reach low values of $\chi^2_{k,j}$ and obtain a measurement of $\tau$ and $C_{F,k}$.

The second algorithm consists of calculating the $K_{k,j}$ ratios over a regular grid of mass thickness and composition, in order to obtain a discrete function $K_{k,j}(\tau^m, C_k^n, C_j^p)$; then this function is interpolated cubically to retrieve the values of mass thickness and composition which minimize the $\chi^2_{k,j}$. The first algorithm is more useful for a fast calculation of film mass thickness and composition in a standard single point measurement, while the second algorithm is necessary when we want to obtain a two-dimensional mass thickness and composition mapping, as the case shown in Section 4.3.

Finally, for the spatial distribution problem, we can solve equation (15) in parallel to equation (6) and (11) with the same numerical method described for equation (6), namely the upwind Euler method for the advection term and the centered second finite difference method for the diffusion term; the boundary conditions are straightforward and the initial condition on $n$ is given by the electron beam radial distribution. Thanks to this step it is possible to numerically retrieve the X-rays generation radial distribution function $\psi(\rho r)$ with equation (16) and to predict the radial resolution, as shown in Fig. 2b.

We implemented the whole model, consisting in the electron transport solution, in the $\phi(\rho z)$ and $\psi(\rho z)$ functions prediction and in the algorithm for the evaluation of film mass thickness and composition of film-substrate systems, in a MATLAB® application, which requires a low calculation time (a single measurement calculation runs in few minutes in a standard computer), called "EDs for areal Density & composItion Evaluation" (EDDIE).

## 3. Model physical parameters

The models described in Section 2 rely on a number of different physical parameters and, in order to obtain enough accurate outputs, namely the mass thickness and composition, they must be calculated with high precision models or database. In this section we describe the literature works that we have chosen to rely on, while in Section 4.1 we describe how the inaccuracy of these parameters reflects on the inaccuracy of the outputs with an uncertainty and sensitivity analysis.

### 3.1. Electron elastic scattering

As described in the Section 2, one of the more important physical input that enables to describe the electron multiple elastic scattering process is the transport mean free path, which appear in equation (6).

We rely on the calculations based on the solution of the Dirac equation, in the approximation of 'static field', which means that the atomic electron density has spherical symmetry, and the differential cross section is evaluated with the relativistic Dirac-Hartree-Fock potential, that is considered the more reliable model for the atomic potential [30,31]. The static field approximation is considered very reliable in this case, because the momentum transfer is higher for the collisions with the close bounded electrons; in this way, the solid state potential, which greatly varies from one material to another, can be neglected and the multiple scattering phenomenon is independent from the aggregation state.

The values of the transport mean free path calculated from the Dirac-Hartree-Fock differential cross section are tabulated in a recent NIST database [32], that covers the primary electron energy range of $0.05 - 300\ keV$.

### 3.2. Electron inelastic scattering

The other crucial aspect of the model is the electron energy loss described by equation (11); so it is of great importance to have a reliable model for the differential inelastic cross section $\partial \sigma_i(E, W)/\partial W$.

Some analytical model for the inelastic cross section exist, but only for quite simple systems, like Hydrogen or free electron gas [33], and are nevertheless complex to calculate; thus, it seems to us reasonable to make use of a semiempirical model which combine a sufficient grade of accuracy with the possibility to obtain the differential inelastic cross section for any kind of material with low calculation time [34].

### 3.3. X-rays generation, attenuation and detection

Finally, it is fundamental for the calculation of X-ray generation in the sample, to know the electron microscopic ionization cross section $\sigma_{ion}(E)$ and the fluorescence yield factor $\omega$, which appear in equation (2); and it is of equal importance to know the X-rays mass attenuation coefficient $\mu/\rho$ and the detection efficiency $\varepsilon$, present in equation (1). We used a recent analytical formula for the ionization cross section [35] that approximate with 1% of error the theoretical ionization cross section calculated from the relativistic distorted-wave Born approximation (DWBA) which consistently account the effect of distortion of the projectile wavefunctions caused by the electrostatic atom field and the exchange effects which arise from the indistinguishability of the projectile and the target electrons. The resulting ionization cross sections have been compared to available experimental data, to other theoretical calculations and to empirical and semi-empirical formulas, showing that the DWBA provides a better description of recent measurements.

The fluorescence yield ω, defined as the probability that an ionized atom emits a characteristic X-ray, is given by empirical fits of experimental data as a function of the atomic number and the ionized shell. Some reliable databases exist that collect all these data [36,37], and we decided to use the most recent one. Nevertheless, some uncertainties still exists for low atomic number elements because the few experimental data obtained in that range quite differ from the tabulated data of the recent databases [38,39]; we noticed that for $Z \leq 8$ the older database values for fluorescence yield are more consistent with experimental data, for this reason we decided to integrate these values in the more recent database.

The X-rays mass attenuation coefficients $\mu/\rho$ are tabulated for all elements over the energy range $0.05 - 30\ keV$ [40] and their values are considered very reliable because they are based on a large quantity of experimental data and established theoretical calculations.

The X-rays detection efficiency $\varepsilon$ depends on the intrinsic detector energy response and the detector window transmittance; these quantities depend on the characteristic X-ray energy and nominal values are given by the detector manufacturer. Nevertheless, these values vary for each detector and they can also change with the aging of the instrument; for this reason some uncertainty should be expected on this parameter. In order to reduce this uncertainty some methods could be exploited in the future: the true detector efficiency curve could be retrieved by measuring the EDS spectrum of an engineered reference sample at a fixed accelerating voltage [41,42].

### 4. Results and discussion

The model for the film mass thickness and composition evaluation, explained in Section 2 and 3, is subjected to an uncertainty and sensitivity analysis, in the first part of this section, with the objective of estimating the $\tau$ and $C_{F,k}$ measurements error bars and to address the physical parameters which contribute more to the measurement uncertainty, in order to pave the way to further improvements.

In the second part the new technique is experimentally validated by the comparison of measurements with other standard and reliable techniques. We prepared various film-substrate samples and we measured the film mass thickness and compositions with standard techniques, namely the EDS for the composition measurements (carried out at $5\ kV$) and the X-Ray Reflectometry (XRR) and standard weighting techniques for the mass thickness evaluation (see Section 6 for a description of the deposition and characterization techniques). In particular the XRR is a standard technique which enables to calculate with high accuracy the electronic density and the thickness of planar thin films and it consists in the collection of monochromatic X-rays reflected by the multilayer sample.

The samples film and substrate compositions were chosen in order to cover a wide range of atomic numbers ($Z_F = 6, 7, 8, 74; Z_S = 14, 42$): Sample 1 is composed of a compact metallic, $64\ nm$ thick, Tungsten film and a metallic Molybdenum substrate, Sample 2 is composed of a compact, $99\ nm$ thick, Carbon film and a Silicon wafer substrate, Sample 3 is an amorphous, $102\ nm$ thick, Tungsten film with Oxygen inclusions on a Molybdenum substrate and Sample 4 is a porous amorphous, $135\ nm$ thick, Tungsten, Oxygen and Nitrogen based film on a Molybdenum substrate.

$K_{k,j}$ ratios were calculated from EDS measurements of each sample at different accelerating voltages; the measurement conditions were fixed to $120\ s$ acquisition time and we selected the $K$ lines for Carbon, Oxygen, Nitrogen and Silicon, while we used the sum of $L_\alpha$ and $L_\beta$ lines for Molybdenum and $M_\alpha$ line for Tungsten.

Finally, we give in Section 4.3, a sample application of the technique: the mapping of mass thickness and composition of a nanostructured ultra-low density and high roughness film, which is difficult to characterize with standard techniques.

**Table 1** Results of the Monte Carlo uncertainty and sensitivity analysis ($N_{it} = 200$) carried out with inputs from Sample 1, with accelerating voltage equal to $10, 20, 30\ kV$. On the last column the assumed relative standard deviations on the parameters are shown.

| Accelerating Voltage $[kV]$ | 10 | 20 | 30 | |
|---|---|---|---|---|
| Output std. deviation $[\mu g/cm^2]$ | 14.9 | 17.5 | 18.5 | |
| Model linearity | 1.07 | 0.98 | 1.04 | |
| **Variance Decomposition** | | | | **Assumed Relative Std Deviation** |
| Elastic scattering CS | 0.01 | 0.03 | 0.00 | 0.10 |
| Inelastic scattering CS | 0.50 | 0.15 | 0.11 | 0.20 |
| Ionization CS | 0.39 | 0.61 | 0.58 | 0.10 |
| Fluorescence yield | 0.03 | 0.14 | 0.11 | 0.05 |
| Detector efficiency | 0.06 | 0.02 | 0.13 | 0.05 |
| Photon attenuation | 0.01 | 0.04 | 0.07 | 0.025 |
| Poisson uncertainty | 0.00 | 0.01 | 0.00 | 0.01 |

**Table 2** Results of the partial derivative variance decomposition analysis, carried out with inputs from Sample 2.

| Accelerating Voltage $[kV]$ | 4 | 5 | 6 | 8 | 10 | 12 | 14 | 16 |
|---|---|---|---|---|---|---|---|---|
| Output std. deviation $[\mu g/cm^2]$ | 2.1 | 2.5 | 2.8 | 3.7 | 4.1 | 4.8 | 4.5 | 5.9 |
| **Variance Decomposition** | | | | | | | | |
| Elastic scattering CS | 0.00 | 0.00 | 0.00 | 0.00 | 0.00 | 0.00 | 0.01 | 0.05 |
| Inelastic scattering CS | 0.77 | 0.58 | 0.50 | 0.37 | 0.42 | 0.40 | 0.33 | 0.11 |
| Ionization CS | 0.14 | 0.23 | 0.31 | 0.43 | 0.35 | 0.39 | 0.37 | 0.35 |
| Fluorescence yield | 0.04 | 0.07 | 0.09 | 0.09 | 0.10 | 0.09 | 0.11 | 0.18 |
| Detector efficiency | 0.04 | 0.07 | 0.09 | 0.09 | 0.10 | 0.09 | 0.11 | 0.18 |
| Photon attenuation | 0.00 | 0.02 | 0.01 | 0.01 | 0.01 | 0.01 | 0.05 | 0.11 |
| Poisson uncertainty | 0.00 | 0.02 | 0.01 | 0.01 | 0.02 | 0.02 | 0.02 | 0.02 |

### 4.1. Uncertainty and sensitivity analysis

The technique uncertainty with respect to the physical parameters errors was investigated through an uncertainty and sensitivity analysis on the underlyng model.

Firstly we carry out the analysis through a Monte Carlo method [43], using as inputs the experimental $\widetilde{K_{k,J}}$ ratios measured from Sample 1 at the accelerating voltages of $10, 20, 30\ kV$. We assume a gaussian distribution function on physical parameters errors, with reasonable standard deviation values assumed from the relative cited literature works and summarized in Table 1, and a Poisson distribution on the $\widetilde{K_{k,J}}$ ratios error; then we extract at each Monte Carlo iteration a different value for each physical parameter following its error distribution, and we obtain different $\tau$ and $C_{F,k}$ values at each iteration. With a sufficiently high number of iterations (in the shown case $N_{it} = 200$) we achieve to find an uncertainty distribution on the output values and consequently we can estimate the standard deviation, namely the error bar.

From the results of Table 1, it should be noted that higher accelerating voltage measurements intrinsically have higher error values. Moreover, we carry out the output variance decomposition in order to determine which physical parameters contribute more to the output uncertainty. For example, in the case of Table 1, the parameters that contribute more to the mass thickness error are the inelastic scattering cross section, which decreases with voltage, the ionization cross section and the fluorescence yield, which increase with accelerating voltage; thus, to further improve the model, the accuracy of these three parameters should be increased.

Moreover the Monte Carlo sensitivity analysis enables to estimate the linearity of the model with respect to the errors; in the above mentioned case, we see that this factor is always near 1 and we can reasonably conclude that the model error behaviour is quite linear. This useful result can be exploited to simplify the uncertainty analysis, in fact in the linear error models the partial derivative analysis can be carried out, which is simpler and faster to perform compared to the Monte Carlo method. Thus, the error bars reported in Section 2.2 measurements are obtained with this method.

In order to confirm the results of the Monte Carlo variance decomposition we carry out the partial derivative variance decomposition also on Sample 2. The results, summarized in Table 2, confirm that the errors increase with the measurement accelerating voltage; in addition it should be noted that also in this case the parameters which are more significant for the error generation are the inelastic cross section, the ionization cross section, the detector efficiency and the fluorescence yield. In addition, we see a strong correlation of the increasing uncertainty with the ionization cross section, as the Monte Carlo analysis highlights for the case of Sample 1. For this reason we expect that increasing, with new measurements and better models, the accuracy of the ionization cross section will result in a strong improvement of the new technique.

### *4.2. Benchmarks*

As described in Section 2, the new method, in contrast to all the EMPA-related literature, overcomes the need of a reference sample, once the substrate composition is known. All the measurements are consequently taken onto only the analysed samples, reducing automatically the number of EDS measurements by a factor 2.

In order to validate the technique reliability we made EDS measurements at many values of accelerating voltage for each sample and we collected the results in Fig. 3, 4, 5. For the Samples 1, 2, 3 (Fig. 3, 4) we used the XRR technique measurements as a benchmark thanks to its high precision and the data reported on the abscissa is the mass thickness, while for Sample 4 we used a standard weighting procedure, because the high roughness of this sample prevents the use of XRR, and the reported data in the abscissa is the density; because the new technique enables to retrieve the mass thickness, the density was calculated by $\rho = \tau/t$, where the film thickness $t$ was measured with cross section SEM image.

All the measured samples prove that the new method mass thickness measurements agree with the benchmarks inside the error bars, calculated with the uncertainty analysis. We see that at low voltages measurements there is a correlation between the mass thickness and the accelerating voltage and we believe that it is caused by the effect of the inaccuracy of inelastic cross sections at low energy values; this point is justified by the fact that at low voltages the sensitivity analysis shows a prevailing role in error generation by this physical parameter (see Table 1 and 2). Nevertheless all the measurement fluctuations with respect to the accelerating voltage are all less then the error bars; accordingly, it is reasonable to state that the new technique can be used, at the limit, with only one accelerating voltage measurement, as opposed to the most existing commercial EMPA software which need many voltages to obtain accurate results.

The mass thickness errors, calculated by the uncertainty analysis, show lower absolute values for the case of Sample 2 with respect to the other ones. This effect is probably caused by lower values of accelerating voltages used for Sample 2 characterization (as explained in Section 4.1, the error slowly increases with the accelerating voltages), and by the lower mass thickness of the film (lower by a $1/5$ factor with respect to the other samples). In all the cases, the mass thickness errors are higher by at least of one order of magnitude with respect to XRR errors, but it should be taken into account that the method does not suffer from limitations due to film roughness, as for the case of Sample 4 which could not be characterized by XRR; in addition, our technique enables to retrieve at the same time the film composition, which is not determined by XRR.

Moreover, we point out that the error bars values lie in the range $2 - 20 \; \mu g/\text{cm}^2$, in all the shown cases. This value is comparable to the resolution of nuclear standard techniques ($\sim 10 \; \mu g/\text{cm}^2$) such as Rutherford Backscattering Spectrometry (RBS) [44] and Time Of Flight – Elastic Recoil Detection Analysis (TOF-ERDA) [45].

Finally, it should be noted that the composition measurements of Sample 3 and 4 (Fig. 4 and 5) are in agreement with the standard EDS composition measurement within few percentage points. Nevertheless, there are some little deviations from the benchmark for higher accelerating voltage values; the variance decomposition analysis (data not shown) indicates that, also in this case, the ionization cross section increases with the accelerating voltage and it is responsible for about the 50% of the error. In addition, we observe that the errors are higher for the Sample 4 with respect to Sample 3, probably for the higher number of elements in that film. Thus, we believe that the knowledge of the ionization cross section parameter should be enhanced in order to finely characterize the composition of films with a large number of elements.

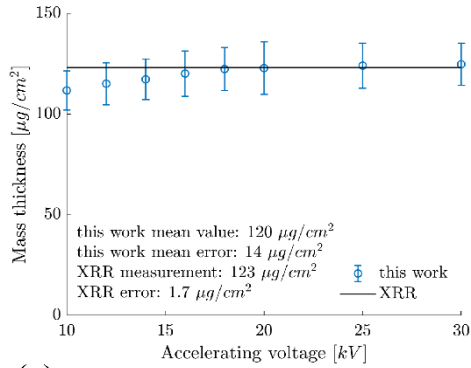 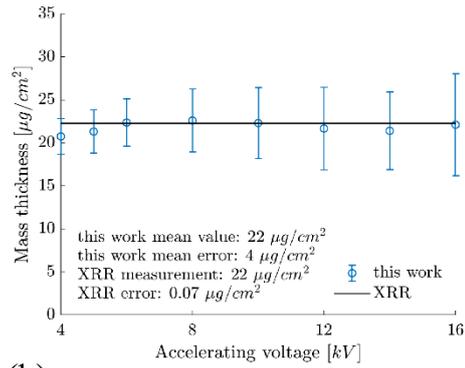

(a) (b)

*Fig. 3. Figure (a) shows the mass thickness of Sample 1 (Tungsten, 64 nm thick, film onto Molybdenum substrate), measured by our method (points in blue) at different accelerating voltages, and by the XRR (line in black). Figure (b) shows the mass thickness of Sample 2 (Carbon, 99 nm thick, film onto Silicon substrate), measured by our method (points in blue) at different accelerating voltages, and by the XRR (line in black). On the graphs the mean value obtained by our method, the mean value of its error bars, the value measured by XRR and its error are displayed.*

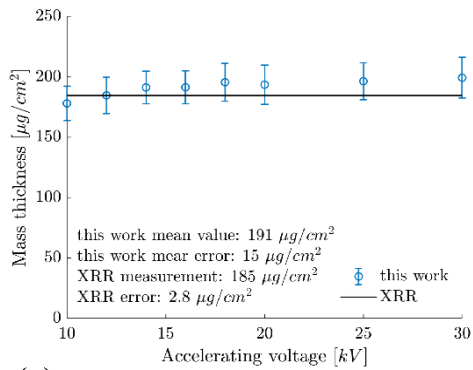 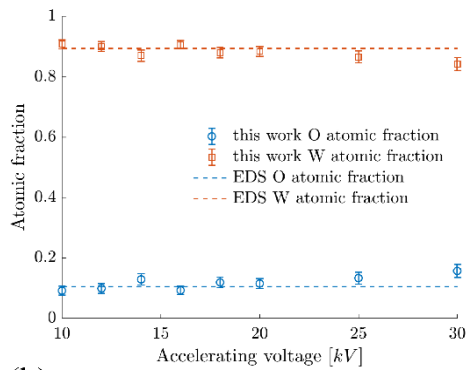

(a) (b)

*Fig. 4. Figure (a) shows the mass thickness and figure (b) the composition of Sample 3 (Tungsten with Oxygen inclusions, 102 nm thick, film onto Molybdenum substrate), measured by our method (points in blue and orange) at different accelerating voltages, and by the XRR (line in black) and EDS (blue and orange dashed lines). On the figure (a) the mean value obtained by our method, the mean value of its error bars, the value measured by XRR and its error are displayed.*

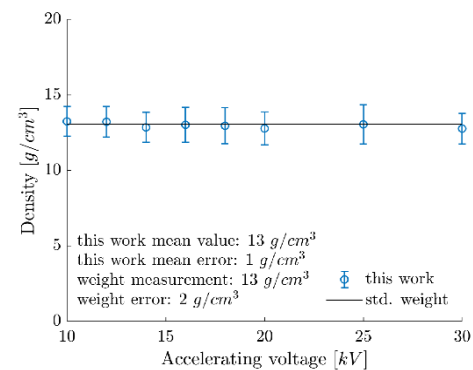 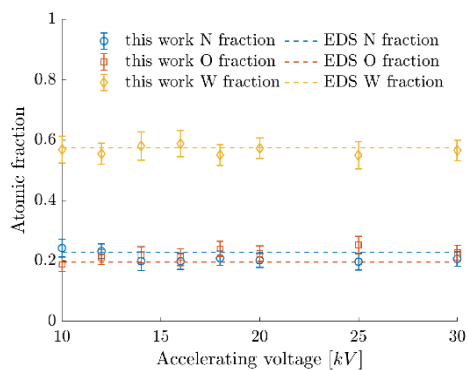

(a) (b)

*Fig. 5. Figure (1) shows the density and figure (b) the composition of Sample 4 (Tungsten with Nitrogen and Oxygen inclusions, 135 nm thick, film onto Molybdenum substrate), measured by our method (points in blue, orange and yellow) at different accelerating voltages, and by the standard weight measurement (line in black) and EDS (blue, orange and yellow dashed lines). On the figure (a) the mean value obtained by our method, the mean value of its error bars, the value measured by XRR and its error are displayed.*

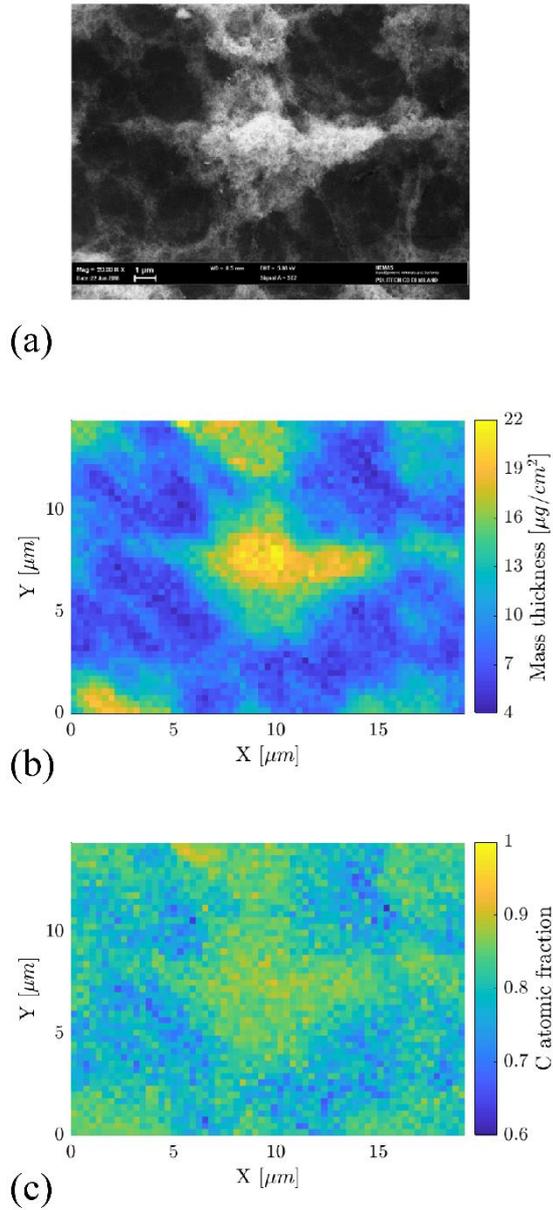

*Fig. 6. Figure (a) shows a SEM image of a Carbon foam film; figure (b) shows the film mass thickness map and figure (c) the Carbon atomic fraction map retrieved with the EDDIE software in the measurement condition of 5 kV accelerating voltage and 300 nm pixel dimension. The spatial resolution is sufficient to highlight the high roughness feature of the film.*

*4.3. Mass thickness and composition mapping: an example*

In this section we give an example application of the technique explained in Section 2 and 3: the mapping of mass thickness and composition of a nanostructured ultra-low density Carbon and Oxygen based film (called Carbon foam [46]). The analysed film was deposited onto a Silicon substrate and has a mean density near $20\ mg/cm^3$ and a mean thickness of 5 $\mu m$ with a very high roughness ($\pm$ 3 $\mu m$ over the $\mu m$ lateral scale), which make this kind of material very difficult to characterize with standard techniques, such as XRR or high sensitivity balances.

The capability of measuring EDS map from a surface over a grid was exploited to retrieve two-dimensional profiles of $K_{k,j}$ ratios with a fixed accelerating voltage value (5 $kV$), which are used, through our method, to retrieve a film mass thickness and composition maps. Fig. 6 shows the mass thickness and composition measurements in comparison with the relative SEM image; it should be observed that the method enables to obtain a mass thickness map (with pixel dimension equal to 300 $nm$) which is in agreement with the qualitative information given by the electron microscope image. The radial resolution $\widetilde{\rho r}$, estimated by the radial distribution equation (16), with beam diameter equal to 10 $nm$, is about $470\ nm$, which is comparable to the pixel dimension; as the mass thickness map highlights the film roughness features, we can consequently state that the radial resolution estimation is in agreement with the experimental data. In addition, we can retrospectively verify the hypothesis of slowly varying mass thickness, which is necessary for the exploitation of the method: $\tilde{r}/t \cdot \partial t/\partial r \sim 3\ \mu m/5\ \mu m \cdot 470\ nm/5\ \mu m = 0.06 \ll 1$ and, consequently, the hypothesis is satisfied.

Finally, we point out that, in general, the radial resolution depends largely on the electron initial energy and the film density; with proper conditions, as low accelerating voltage ($< 10\ kV$) and bulk density films ($> 1\ g/cm^3$) the radial resolution can reach much lower values, down to values limited only by the electron beam diameter.

### 5. Conclusions

In conclusion, we have described a new method for the evaluation of thin films mass thickness and composition from EDS data, which is very appealing because it is non-destructive, it needs a common experimental apparatus, it has a high spatial resolution, it is not affected by film roughness and it does not need a reference sample. The method relies on a numerical solution of a simplification of the Boltzmann transport equation for electrons, based on reasonable assumptions. This theoretical approach enables us to calculate the $\phi(\rho z)$ function, which describes the distribution of X-ray generation with depth of the sample, also in a film-substrate geometry, with high accuracy. Thanks to this fact this method for the evaluation of mass thickness and composition does not need a reference sample and multiple voltages measurements, and the implemented software, called EDDIE, can run in standard computers in a few minutes. We also point out that the derived method could be implemented in the future also in more complex geometries, for example in more than two layers samples, or with depth-dependent composition.

The method was validated with benchmarks characterized by standard techniques and an uncertainty and sensitivity analysis was carried out in order to estimate the errors relative to the mass thickness and composition. The analysis highlighted that the mass thickness measurement errors lie in the range $2 - 20\ \mu g/cm^2$ which is comparable to other nuclear standard techniques (such as RBS and ERDA), while the sensitivity analysis indicated that the method accuracy could be strongly enhanced by increasing the accuracy of the electron ionization cross sections and the electron inelastic cross sections.

Finally, we have shown an example application of the method: we obtained accurate mass thickness and composition maps of an ultra-low density nanostructured film, with a spatial resolution properly predicted from our model. Thanks to the demonstrated capabilities of the technique and to its high accuracy, we believe that this new method should play a role as a new standard technique for mass thickness and composition determination.

### 6. Materials & methods: deposition and characterization techniques

The samples were produced by the deposition of thin films by the Pulsed Laser Deposition technique (PLD). For Samples 1, 3, 4 we used the second harmonic, $\lambda = 532\ nm$, pulse of a Nd:YAG laser, duration $5-7\ ns$ and repetition rate $10\ Hz$, while for Sample 2 we used the fundamental, $\lambda = 1054\ nm$; the beam was directed on a $2\ in$ Tungsten target, for Samples 1, 3, 4, and pyrolytic graphite target, for Sample 2, with 45° angle of incidence. The ablated species expanded from the target to a Silicon substrate distant $5\ cm$ from the target, for Sample 2, and a Molybdenum substrate distant $7\ cm$ from the target, for Samples 1, 3, 4, kept at room temperature, inside a home made vacuum chamber evacuated by a primary scroll pump and a turbo-molecular pump, reaching a base pressure of $10^{-3}\ Pa$. The pulse energy and spot were varied for the different samples and the fluence on target was fixed at $11.3\ J/cm^2$ for Samples 1, 3, 4 and $1.2\ J/cm^2$ for Samples 2. We used two different gasses to fill the chamber, $N_2$ and He; in particular, Sample 1 was produced in vacuum ($10^{-2}\ Pa$), Sample 2 was produced with $20\ Pa$ of He, Sample 3 with $70\ Pa$ of He and Sample 4 with $5\ Pa$ of $N_2$.

The ultra-low density Carbon foam film was deposited by the femtoseconds PLD technique which exploits the fundamental frequency, $\lambda = 800\ nm$, of a Ti:Sa laser, duration $80\ fs$ and repetition rate $1\ kHz$. The laser pulses ablated a pyrolytic graphite target with 45° angle of incidence and a fluence of $0.16\ J/cm^2$ and the ablated species were collected onto a Silicon substrate, distant $7\ cm$ from the target, in an Ar gas atmosphere at $100\ Pa$.

The SEM cross section measurement and the composition EDS measurements were taken with a Zeiss Supra 40 field emission SEM in combination with an Oxford Instruments Si(Li) detector.

For the mass thickness benchmark measurements of Sample 1, 2, 3 we exploited the X-Ray Reflectometry technique (XRR), which consists of the collection of the reflected X-rays (reflected vector) from the sample originated by a monochromated X-ray beam (incident vector). Experimentally, this is achieved as follows. X-rays generated from a Cu source ($40\ kV$, $0.9\ mA$) are monochromated by a parabolic mirror to obtain a parallel X-ray beam of wavelength $\lambda =$

$0.154\ nm$ (Cu $K_\alpha$). The incident X-ray beam is sized down to $6\ mm\ \times\ 0.1\ mm$, with the larger value in the transverse direction with respect to the beam axis, by slits to minimize spill-off from sample edges. The reflected X-rays are collected with a point Na:Tl solid state scintillator, with acceptance slits of $0.2\ mm$, to minimize the collection of scattered X-rays in non-reflection condition, positioned at a distance of $40\ cm$ from the sample stage. Measurements are collected in a symmetric $\Omega = 2\Theta$ geometry, where $\Omega$ is the angle between the incident X-rays and the sample plane and $\Theta$ is the angle between the sample plane and the detector, from $\Omega = 2\Theta = 0°$ to $\Omega = 4°, 2\Theta = 8°$ in steps $\Delta\Omega = 0.01°$ and $\Delta(2\Theta) = 0.02°$ with $2\ s$ or $10\ s$ collection time for each point. Data are fitting with MAUD software, which implements an algorithm based on matrix formalism model corrected by a Croce-Nevot factor. In particular, the electron density $\rho_e$ is obtained from the critical angle $\Theta_c$, related to the critical vector $q_c$. For Cu $K_\alpha$ X-ray emission $q_c$ and $\rho_e$ are linked by the equation $q_c\ (\text{Å}^{-1}) = 0.0375\ \sqrt{\rho_e\ (\text{Å}^{-3})}$ [47-51].

For the density benchmark measurement of Sample 4 a standard weighting procedure was made though a high precision balance and the density was calculated dividing the mass from the film volume.

## Acknowledgements

This project has received funding from the European Research Council (ERC) under the European Union's Horizon 2020 research and innovation programme (ENSURE grant agreement No. 647554).

## Competing interests

The authors declare no conflict of interest.

## Author contributions

A.P. developed and implemented the model, analysed the data, wrote the manuscript and produced the tables and figures. A.M. and D.D. produced the analysed samples. A.L. characterized the samples with XRR and wrote the relative Methods section. M.P. conceived the project and supervised all the activities. All authors reviewed the manuscript.

## Data availability

The raw EDS data required to reproduce these findings are available on request. The EDDIE code required to reproduce these findings is available to download from https://data.mendeley.com/datasets/ytmyshtn66/draft?a=ee3dbded-7061-4217-a99b-578ade5c62ed.